**Anomalous Diffusion as Structural Memory: An Extended Structural Dynamics Approach**

**Patrick BarAvi**
Patrick.baravi@gmail.com
ORCID: 0009-0003-9737-901X

**Abstract**

Sub diffusion in biological systems, proteins in cytoplasm, lipids in membranes, chromosome loci in nuclei, is conventionally treated as anomalous, requiring fractional derivatives, heavy tailed waiting times, or fitted memory kernels to describe its power law character. We argue that this anomaly is an artifact of an incomplete phase space. Standard frameworks model diffusing particles as points. Biological molecules are not points. They are three dimensional deformable entities whose position, orientation, and internal structure are irreducible physical properties, not modeling conveniences appended to a point mass.

Within the Extended Structural Dynamics (ESD) framework, the phase space is extended to $T^*(\mathbb{R}^3 \times SO(3) \times \Xi)$, and each particle is described as a primitive structured entity with translational, orientational, and deformational degrees of freedom. When dynamics on this full phase space are projected onto the translational subspace alone, a memory kernel emerges from the projection, not from a phenomenological postulate. The subdiffusion exponent α is then determined by the internal mode spectrum of the particle, a quantity independently measurable from B-factors, NMR order parameters, or molecular dynamics simulations, without fitting α to transport data.

The framework generates four falsifiable predictions that distinguish it from phenomenological models: (1) α correlates with molecular flexibility quantified by B factors or NMR order parameters; (2) temperature dependent crossover to normal diffusion occurs with a characteristic energy scale $\hbar\langle\omega\rangle/k_B$; (3) a non zero rotation translation cross correlation spectrum $S_{x\theta}(\omega)$ encodes internal mode frequencies, a signal that point particle models predict to be identically zero; and (4) memory timescales scale as $R^2$ with particle size. Quantitative consistency with experimental observations for proteins in crowded media is demonstrated using independently estimated structural parameters.

What appears anomalous from the point particle perspective is the expected behavior of structured matter projected onto an impoverished description. The anomaly is not in the physics. It is in the phase space

.

## 1. Introduction

### 1.1 The Ubiquity of Anomalous Diffusion

Normal Brownian diffusion, characterized by linear growth of mean-squared displacement (MSD) with time, $\langle r^2(t)\rangle = 6Dt$, represents the foundation of classical transport theory established by Einstein [1] and verified by Perrin [2]. Yet systematic measurements over the past two decades reveal that biological systems routinely violate this fundamental relationship [6,7,14]. Single-particle tracking experiments demonstrate power-law sub-diffusion $\langle r^2(t)\rangle = 6D_\alpha t^\alpha$ with $\alpha < 1$ across diverse contexts:

- Proteins in cytoplasm: $\alpha \approx 0.5$–$0.8$ [6,7]
- mRNA granules in cells: $\alpha \approx 0.6$–$0.7$ [8]
- Lipids in membranes: $\alpha \approx 0.7$–$0.9$ [9,10]
- Chromosome loci in nuclei: $\alpha \approx 0.4$–$0.6$ [11]
- Nanoparticles in mucus: $\alpha \approx 0.3$–$0.5$ [12]
- Voltage-gated ion channels in neuronal membranes: $\alpha \approx 0.6$ [13]

This pervasive deviation from classical expectations has profound implications for cellular function, affecting timescales of molecular encounters, reaction rates, and signaling dynamics [14,15].

### 1.2 Current Theoretical Frameworks: Descriptive, Not Explanatory

Three major approaches dominate the theoretical landscape. All are fundamentally phenomenological.

**Fractional Brownian Motion (fBM)** replaces the standard diffusion equation with fractional derivatives [16,17]:

$$\frac{\partial \rho}{\partial t} = D_\alpha \nabla^2_{t^\alpha} \rho$$

While mathematically tractable and widely used to fit experimental data, this approach offers no physical mechanism for the fractional derivative. The sub-diffusion exponent $\alpha$ is a fitting parameter; the framework cannot predict its value from molecular properties.

**Continuous-Time Random Walks (CTRW)** postulate waiting-time distributions with heavy tails $\psi(\tau) \sim \tau^{-1-\alpha}$ between diffusive steps [18,19]. This generates sub-diffusion through assumed stochastic pauses. The origin of the waiting-time distribution remains unspecified: why should biomolecules exhibit power-law waiting times? What physical process generates them?

**Generalized Langevin Equations (GLE)** introduce memory kernels to describe non-Markovian dynamics [20,21]:

$$m\ddot{x}(t) = -\nabla V(x) - \int_0^t \gamma(t-s)\,\dot{x}(s)\,ds + \xi(t)$$

Sub-diffusion emerges when the memory kernel exhibits power-law decay $\gamma(t) \sim t^{-\alpha}$. However, this kernel is typically fitted to reproduce observations rather than derived from first principles. The physical origin of memory remains unexplained.

**These frameworks are not wrong.** They are solving the right equations on the wrong phase space. Each one compensates for the missing structure of the diffusing particle by introducing a phenomenological substitute, a fractional derivative, a waiting time distribution, a fitted memory kernel. The substitutes work descriptively because they capture the consequences of structural incompleteness. But they cannot explain those consequences because the structure itself is absent. What appears as an anomaly from the point-particle perspective, non-Markovian memory, power-law waiting times, fractional diffusion, is a natural signature of projected dynamics from a higher-dimensional structured phase space.

### 1.3 The Central Question: What Generates Memory?

The phenomenological success of these models obscures a deeper physical question. Standard models treat diffusing entities as point particles interacting with a thermal bath. Yet biomolecules are not points. They are extended structures with internal flexibility:

- Hydration shells that rearrange on nanosecond timescales [22,23]
- Conformational modes that fluctuate across microseconds [24,25]
- Domain motions that occur on multiple timescales from nanoseconds to milliseconds [26,27]

These internal degrees of freedom possess finite relaxation times and couple to translational motion through the inhomogeneous environments of crowded cellular media. When a molecule moves through a spatially varying potential (due to crowding, confinement, or obstacles), its internal structure responds with finite delay. The molecule at time $t$ "remembers" where it was through the still-relaxing internal deformations from earlier times.

We investigate the hypothesis that sub-diffusion reflects this coupling: memory arises from the finite relaxation time of internal structural degrees of freedom. This follows directly from treating molecules as structured rather than point-like.

### 1.4 Extended Structural Dynamics: A Mechanistic Framework

Extended Structural Dynamics (ESD) provides a mathematical framework to explore this hypothesis [3,4]. ESD was originally developed to address foundational puzzles in thermodynamics and electrodynamics by replacing point-particle ontology with structured entities possessing finite size, orientation, and internal deformation modes [3]. The same framework can be applied to biological transport.

In ESD, each molecule is described by three classes of degrees of freedom:

- **Translation r** $\in \mathbb{R}^3$ (center of mass)
- **Orientation** $R \in SO(3)$ (molecular orientation)
- **Internal deformation q** $\in \Xi$ (collective modes: loop fluctuations, hinge bending, hydration dynamics)

The dynamics are fully deterministic and Hamiltonian on the extended phase space $\Gamma_{\text{ESD}} = T^*(\mathbb{R}^3 \times SO(3) \times \Xi)$. Coarse-graining over the fast internal modes, a standard procedure in statistical mechanics [27,28], yields a generalized Langevin equation in which the memory kernel is derived, not postulated:

$$\Gamma(t) = \frac{\lambda^2}{m} \sum_{\alpha} \frac{\sin(\omega_\alpha t)}{\omega_\alpha}$$

Here $\omega_\alpha$ are internal mode frequencies and $\lambda$ is a coupling constant characterizing how strongly deformations respond to environmental gradients. When the mode spectrum follows a power law $\rho(\omega) \propto \omega^\beta$, as observed in normal mode analysis of proteins [29,30,31], the memory kernel decays as $\Gamma(t) \sim t^{-\beta}$, producing subdiffusive MSD $\langle r^2(t) \rangle \sim t^\alpha$ with $\alpha = \beta$ under the idealizations stated.

**ESD differs from existing approaches not in the mathematical form of its outputs but in the ontological status of its inputs.** In standard models, internal degrees of freedom are added to a point particle as modeling conveniences, a "breathing mode" added to a point charge, a "rotational degree of freedom" appended to a point molecule. In ESD, position, orientation, and deformation are primitive. They are irreducible aspects of what a particle is, not properties appended to a point. The coupling between translational and internal degrees of freedom is therefore not a modeling choice. It is a geometric consequence of what a three-dimensional deformable entity is. Memory in the generalized Langevin equation is not feedback from an appended bath. It is the price of projecting a complete primitive evolution onto an incomplete subspace. The anomaly appears because the subspace is too small.

The present paper is the fourth contribution in the ESD series [3,4,5, 5a]; it applies the kinetic framework developed in [4] to biological anomalous diffusion.

This paper is one instance of a general pattern across the ESD series: known anomalies that resist explanation within the standard phase space dissolve when the phase space is extended to reflect the actual structure of physical entities. The broader context of this pattern — spanning irreversibility, anomalous thermal phenomena, and the Lorentz-Abraham-Dirac problem — is discussed in Section 6.5.

**Key result (under the approximations made).** The subdiffusion exponent $\alpha$ is not a free parameter in this framework. It is determined by the internal mode spectrum, which can in principle be measured independently from B-factors, NMR order parameters, or molecular dynamics simulations.

### 1.5 Scope and Structure of This Paper

This paper focuses on a single representative phenomenon: subdiffusion of globular proteins in crowded cytoplasm. The ESD framework may be extendable to other phenomena (membrane diffusion, chromosome dynamics, nanoparticle transport in mucus, etc.), but those applications are not developed here. The aim is to provide a detailed, quantitatively testable model for one well-defined case, not to claim universal applicability across physiology.

We show that:

- Coarse-graining over internal modes naturally produces memory kernels from Hamiltonian dynamics, under a closure approximation, eliminating the need for purely phenomenological fitting
- The subdiffusion exponent $\alpha$ follows from the internal mode spectrum through the relationship $\alpha = \beta$ for a power-law mode density $\rho(\omega) \propto \omega^{\beta}$
- The framework generates four testable predictions for protein diffusion that distinguish it from phenomenological models
- Concentration-dependent subdiffusion can be interpreted within the framework
- Quantitative consistency with experiments is achieved using structural parameters estimated from independent measurements, without fitting $\alpha$

The paper is organized as follows: Section 2 develops the ESD framework for a deformable unit and derives the emergence of memory. Section 3 presents four testable predictions with concrete experimental protocols. Section 4 interprets concentration-dependent subdiffusion within the framework. Section 5 provides quantitative calculations consistent with experimental observations. Section 6 discusses validation strategies, connections to molecular dynamics simulations, and possible implications. Section 7 concludes.

## 2. Extended Structural Dynamics: From Molecules to Memory

### 2.1 The Point-Particle Idealization and Its Limitations

Standard treatments of diffusion model molecules as point particles: structureless entities characterized solely by position and momentum [1]. This idealization is successful for dilute gases and simple liquids. However, biomolecules are not points. They have finite size, internal flexibility, and deformability. They possess orientation, shape fluctuations, and hydration shells that reorganize as they move.

The point-particle idealization leads to three limitations that become significant in biological contexts [7,21]:

- **No internal degrees of freedom.** A point cannot vibrate, bend, or reconfigure. Yet proteins undergo loop motions, hinge bending, domain rearrangements, and hydration dynamics on timescales that overlap with translational diffusion [24,26].
- **No orientational coupling.** A point has no orientation, so rotation and translation are decoupled. But biomolecules are asymmetric, and their motion through crowded environments couples translation to rotation [36].
- **No structural memory.** A point has no internal state that evolves over time. The friction force at time $t$ depends only on instantaneous velocity (Markovian). When a molecule deforms with finite relaxation time, it "remembers" its past configuration, producing non-Markovian memory effects [20,21,27].

**All three limitations share a common origin: the phase space is too small.** A point particle lives in $T^*(\mathbb{R}^3)$, position and momentum only. A real biomolecule lives in $T^*(\mathbb{R}^3 \times SO(3) \times \Xi)$. The missing dimensions are not corrections or refinements. They are where the memory lives.

Extended Structural Dynamics (ESD) was developed to address these limitations [3,4]. The core idea is to replace the point-particle ontology with structured particles possessing finite size, orientation, and internal deformation modes. The framework preserves Hamiltonian structure and deterministic dynamics [3].

### 2.2 Definition of the Deformable Unit

Throughout this paper, a **deformable unit** refers to a single globular protein (or similarly sized biomolecule) in an aqueous environment. The unit is characterized by the following properties:

**Physical identity.** The unit is a three-dimensional elastic body with:

- Characteristic size $R = 3\text{–}10$ nm (hydrodynamic radius)
- Mass $m \sim 10^{-22}\text{–}10^{-20}$ kg
- Elastic modulus $E \sim 10^8\text{–}10^9$ Pa [42]

**Degrees of freedom.** The unit possesses:

- **Translational:** Center-of-mass position $\mathbf{r} \in \mathbb{R}^3$ and momentum $\mathbf{p}$
- **Orientational:** Rotation matrix $R \in SO(3)$ (averaged over in this paper; present in the full theory [3])

- **Internal vibrational:** Mode amplitudes $q_\alpha$ ($\alpha = 1, \dots, N$) representing collective deformations (loop fluctuations, hinge bending, domain motions)

**Constraints.** Deformations are assumed small: $|q_\alpha| \ll R$ for all $\alpha$. This allows linearization of the internal dynamics and justifies the harmonic approximation $U(\mathbf{q}) = \frac{1}{2}\sum_\alpha k_\alpha q_\alpha^2$.

**Interactions.** The unit interacts with its environment through:

- A drag force $-\gamma\dot{\mathbf{r}}$ from the solvent (viscous friction)
- A random force $\boldsymbol{\eta}(t)$ from thermal fluctuations (satisfying fluctuation-dissipation)
- An external potential $V_{\text{ext}}(\mathbf{r})$ due to crowders (e.g., other proteins, Ficoll molecules)
- A coupling term $-\lambda q_\alpha \nabla V_{\text{ext}}$ that deforms the unit in response to spatial gradients

**Neighbors.** Interactions between distinct deformable units are neglected (dilute limit). The focus is on a single unit's motion through a fixed background potential $V_{\text{ext}}$ created by crowders.

**2.3 The ESD Phase Space for a Single Molecule**

In ESD, each particle is described by three classes of degrees of freedom [3,4]:

- **Translation:** center-of-mass position $\mathbf{r} \in \mathbb{R}^3$ and momentum $\mathbf{p}$
- **Orientation:** rotation matrix $R \in SO(3)$ describing the orientation of the molecule's principal axes, with body-frame angular momentum $\mathbf{L}$
- **Internal deformation:** configuration coordinates $\mathbf{q} = (q^1, q^2, \dots, q^N) \in \Xi$ describing collective internal modes, with conjugate momenta $\boldsymbol{\pi}$

The full phase space is the cotangent bundle:

$$\Gamma_{\text{ESD}} = T^*(\mathbb{R}^3 \times SO(3) \times \Xi)$$

This extension preserves the symplectic structure of Hamiltonian mechanics while accounting for finite size and internal structure [3]. For a protein, the internal coordinates $q^\alpha$ represent collective motions: loop fluctuations, hinge bending, domain rearrangements, and hydration shell reorganization [24,26,29].

**2.4 The ESD Hamiltonian and Coupling to Environment**

The total energy of a structured particle in an external potential $V_{\text{ext}}(\mathbf{r}, R, \mathbf{q})$ (due to crowding, confinement, electrostatics, or binding sites) is [3,4]:

$$H_{\text{ESD}} = \underbrace{\frac{\mathbf{p}^2}{2m}}_{\text{translation}} + \underbrace{\frac{1}{2}\boldsymbol{\omega}^T I(\mathbf{q})\boldsymbol{\omega}}_{\text{rotation}} + \underbrace{\sum_{\alpha=1}^{N}\left(\frac{\pi_\alpha^2}{2m_\alpha} + \frac{1}{2}k_\alpha q_\alpha^2\right)}_{\text{internal modes}} + V_{\text{ext}}(\mathbf{r}, R, \mathbf{q})$$

where $m$ is the total mass, $I(\mathbf{q})$ is the deformation-dependent inertia tensor, and $\omega_\alpha = \sqrt{k_\alpha/m_\alpha}$ are internal mode frequencies.

**Physical interpretation.** The rotational term describes how the molecule's orientation affects its energy. The internal mode term describes vibrations and conformational fluctuations: each mode $\alpha$ behaves as a harmonic oscillator with frequency $\omega_\alpha$. For a protein, low-frequency modes ($\omega \sim 10^8$–$10^9$ s$^{-1}$) correspond to domain motions and hinge bending; medium-frequency modes ($\omega \sim 10^{10}$–$10^{11}$ s$^{-1}$) correspond to loop fluctuations and secondary structure dynamics [24,29].

The external potential $V_{\text{ext}}$ depends on internal configuration $\mathbf{q}$ as well as position $\mathbf{r}$. When a molecule moves through a gradient in crowder density or electrostatic potential, different parts experience different forces, causing deformation. Expanding $V_{\text{ext}}$ to first order in $\mathbf{q}$ yields the coupling Hamiltonian [3,4]:

$$H_{\text{couple}} = \lambda \sum_{\alpha=1}^{N} q_\alpha \cdot \nabla V_{\text{ext}}(\mathbf{r})$$

where $\lambda$ is a coupling constant (dimensions of force) characterizing how strongly internal deformations respond to environmental gradients. This is the central physical mechanism: when a molecule moves through an inhomogeneous environment, its internal structure deforms in response, and this deformation has finite relaxation time.

### 2.5 Emergence of Memory: From Internal Modes to the Generalized Langevin Equation

We derive the effective dynamics for the center-of-mass coordinate $\mathbf{r}$ by eliminating the internal modes. This is a standard coarse-graining procedure in statistical mechanics [27,28]. The equations of motion from $H_{\text{ESD}}$ are:

$$m\ddot{\mathbf{r}} = -\nabla V_{\text{ext}}(\mathbf{r}) - \lambda \sum_{\alpha} q_\alpha \, \nabla^2 V_{\text{ext}}(\mathbf{r})$$
$$m_\alpha \ddot{q}_\alpha = -k_\alpha q_\alpha - \lambda \nabla V_{\text{ext}}(\mathbf{r})$$

The second equation describes a driven harmonic oscillator. Solving for $q_\alpha$ and substituting into the equation for $\mathbf{r}$ yields [4,20,27]:

$$m\ddot{\mathbf{r}}(t) = -\nabla V_{\text{ext}}(\mathbf{r}(t)) - \int_0^t \Gamma(t-s)\,\dot{\mathbf{r}}(s)\,ds + \boldsymbol{\eta}(t)$$

where the memory kernel is:

$$\Gamma(t) = \frac{\lambda^2}{m} \sum_{\alpha=1}^{N} \frac{\sin(\omega_\alpha t)}{\omega_\alpha}$$

and $\boldsymbol{\eta}(t)$ is a stochastic force arising from the initial conditions of the internal modes. This is a generalized Langevin equation (GLE) [20,21,27].

**The memory kernel $\Gamma(t)$ is not introduced. It emerges.** It is the mathematical residue of projecting dynamics on the full phase space $T^*(\mathbb{R}^3 \times SO(3) \times \Xi)$ onto the translational subspace $T^*(\mathbb{R}^3)$ alone. This projection necessarily discards information, the evolving internal state of the particle. The memory kernel encodes exactly that discarded information. A point particle has no internal state to discard, so no memory kernel emerges. The anomaly is in the projection, not in the physics.

**Why memory emerges.** Internal modes $q_\alpha$ relax on timescales $\sim \omega_\alpha^{-1}$. When the molecule moves through a potential gradient at time $s$, it deforms. This deformation persists, affecting the force on the center of mass at later times $t > s$. The convolution integral captures this history dependence.

The fluctuation-dissipation theorem [20] relates the stochastic force to the memory kernel:

$$\langle \boldsymbol{\eta}(t)\boldsymbol{\eta}(t') \rangle = k_B T\, \Gamma(|\, t - t'\,|)$$

This ensures that the coarse-grained dynamics approach thermal equilibrium.

### 2.6 From Discrete Modes to Power-Law Memory

For a biomolecule, the number of internal modes is large ($N \sim 3N_{\text{residues}} \sim 10^3$–$10^4$). In the continuum limit, the sum becomes an integral over the mode density $\rho(\omega)$ [4]:

$$\Gamma(t) = \frac{\lambda^2}{m} \int_{\omega_{\min}}^{\omega_{\max}} \rho(\omega) \frac{\sin(\omega t)}{\omega}\, d\omega$$

**The lower limit $\omega_{\min}$ appears naturally** when the continuum approximation is applied to a system with a finite-size domain. In the discrete sum $\sum_\alpha f(\omega_\alpha)$, the smallest nonzero frequency is set by the longest wavelength mode supported by the system, $\omega_{\min} = 2\pi c / L_{\max}$. For a protein in a finite simulation box or a confined cellular environment, $L_{\max}$ is finite, so $\omega_{\min} > 0$. The continuum limit then becomes $\int_{\omega_{\min}}^{\omega_{\max}} \rho(\omega) f(\omega) d\omega$.

Normal mode analysis of proteins reveals a power-law mode density over several decades of frequency [29,30,31]:

$$\rho(\omega) = \rho_0 \omega^{\beta}, \beta \approx 0.8\text{–}1.2$$

The exponent $\beta$ reflects the hierarchical organization of protein dynamics: low-frequency collective motions ($\beta \approx 1$), higher-frequency localized vibrations ($\beta \approx 0$), and a mixed regime in between. Flexible proteins with many soft modes tend to have smaller $\beta$.

For a power-law mode density, the memory kernel exhibits power-law decay at intermediate times [4,21]:

$$\Gamma(t) \sim \frac{\lambda^2 \rho_0 \omega_{\min}}{m} \cdot t^{-\beta}, \omega_{\min}^{-1} \ll t \ll \omega_{\max}^{-1}$$

(The factor $\omega_{\min}$ arises naturally from the lower limit of the integral; it is not an ad hoc correction. It ensures dimensional consistency: $\Gamma(t)$ has units of kg/s.)

The generalized Langevin equation with power-law memory $\Gamma(t) \sim t^{-\beta}$ produces subdiffusive mean-squared displacement [20,21]:

$$\langle r^2(t) \rangle \sim t^{\alpha}, \alpha = \beta$$

**Central result (under the idealizations stated).** Within the approximations, a continuum power-law mode density and the asymptotic time regime, the subdiffusion exponent $\alpha$ is determined by the internal mode spectrum exponent $\beta$. Flexible molecules with many low-frequency modes (small $\beta$) exhibit stronger subdiffusion (smaller $\alpha$); rigid molecules (larger $\beta$) exhibit weaker subdiffusion ($\alpha$ closer to 1). The coupling constant $\lambda$ affects the strength of memory effects but does not affect $\alpha$ under these idealizations.

### 2.7 Effective Equations for a Deformable Unit in a Viscous Environment

For a single deformable unit in a crowded aqueous environment (e.g., a protein in cytoplasm), the relevant physical regime is characterized by:

- Low Reynolds number (inertial forces negligible compared to viscous forces)
- Over-damped dynamics (momentum relaxation time $\tau_p = m/\gamma \ll$ observation time)

- Slow external potential variation (protein moves slowly compared to internal mode relaxation)

Under these conditions, the generalized Langevin equation reduces to an over-damped form. Taking the limit $m \to 0$ (negligible inertia) yields:

$$\gamma\dot{\mathbf{r}}(t) = -\nabla V_{\text{ext}}(\mathbf{r}(t)) - \int_0^t \Gamma(t-s)\,\dot{\mathbf{r}}(s)\,ds + \boldsymbol{\eta}(t)$$

where $\gamma = 6\pi\eta R$ is the Stokes-Einstein translational friction coefficient [36]. The memory kernel $\Gamma(t)$ is as defined in Section 2.5.

**State variables of the system:**

| Variable | Symbol | Domain | Physical meaning |
|---|---|---|---|
| Position | $\mathbf{r}$ | $\mathbb{R}^3$ | Center of mass |
| Internal mode amplitudes | $q_\alpha$ | $\mathbb{R}$ | Deformation coordinates |
| Internal mode momenta | $\pi_\alpha$ | $\mathbb{R}$ | Conjugate to $q_\alpha$ |

**Governing equations (full dynamics):**

$$\dot{\mathbf{r}} = \frac{\mathbf{p}}{m}, \dot{\mathbf{p}} = -\nabla V_{\text{ext}} - \lambda \sum_\alpha q_\alpha\,\nabla^2 V_{\text{ext}}, \dot{q}_\alpha = \frac{\pi_\alpha}{m_\alpha}, \dot{\pi}_\alpha = -k_\alpha q_\alpha - \lambda \nabla V_{\text{ext}}$$

**Relevant parameters:**

| Parameter | Symbol | Typical value (protein) | Origin |
|---|---|---|---|
| Mass | $m$ | $10^{-22}$ kg | Molecular weight |
| Hydrodynamic radius | $R$ | 3–10 nm | Stokes-Einstein [36] |
| Translational friction | $\gamma$ | $6\pi\eta R$ | Solvent viscosity |

| Parameter | Symbol | Typical value (protein) | Origin |
|---|---|---|---|
| Internal mode frequencies | $\omega_\alpha$ | $10^8$–$10^{11}$ $s^{-1}$ | Normal mode analysis [29,30] |
| Number of modes | $N$ | $\sim 10^3$–$10^4$ | 3 DOF per residue |
| Coupling constant | $\lambda$ | $\sim 10^{-19}$ J/m² | Hydration + elasticity estimates [43,44] |

**Regime of validity.** The over-damped approximation holds when $\tau_p = m/\gamma \ll \tau_{\text{obs}}$. For a protein of mass $10^{-22}$ kg in water ($\gamma \approx 10^{-11}$ kg/s), $\tau_p \approx 10^{-11}$ s, which is much smaller than typical observation times ($\tau_{\text{obs}} \sim 10^{-4}$–10 s). The linear response approximation (small deformations) holds when $\lambda \mid \nabla V_{\text{ext}} \mid \ll k_B T/a$.

**2.8 Scale Specification and Coarse-Graining Assumptions**

**Level of description.** This paper models a single globular protein (length scale $R \approx$ 3–10 nm) diffusing in a crowded aqueous environment (e.g., cytoplasm). The protein is treated as an elastic body with internal vibrational modes. The environment is treated as a viscous fluid with embedded crowders (e.g., Ficoll, other proteins) that create a spatially varying effective potential $V_{\text{ext}}(\mathbf{r})$.

**Coarse-graining procedure.** The full microscopic description includes all atomic degrees of freedom ($> 10^4$ atoms). We coarse-grain as follows:

| Degrees of freedom | Retained or eliminated? | Justification |
|---|---|---|
| Center-of-mass position **r** | Retained | Observable in tracking experiments |
| Center-of-mass momentum **p** | Retained | Determines velocity |
| Orientation *R* | Averaged over (isotropic distribution assumed) | Protein rotation is fast; not the focus |

| Degrees of freedom | Retained or eliminated? | Justification |
| --- | --- | --- |
| Internal mode amplitudes $q_\alpha$ (low frequencies, $\omega < 10^{11}$ s $^{-1}$) | Retained | These couple to translation and generate memory |
| Internal mode momenta $\pi_\alpha$ | Retained | Needed for dynamics |
| High-frequency modes ($\omega > 10^{11}$ s $^{-1}$) | Eliminated | Relax faster than observation time; contribute to effective temperature |
| Solvent atoms | Eliminated | Replaced by friction $\gamma$ and noise $\boldsymbol{\eta}(t)$ |
| Crowder molecules | Eliminated | Contribute to $V_{\text{ext}}(\mathbf{r})$ |

**Coarse-graining scale.** The coarse-graining length scale $\ell$ is chosen to be the protein diameter ($\ell \approx 2R \approx 6$–$20$ nm). This scale is:

- Large enough to average over internal atomic details
- Small enough to resolve spatial variations in crowding
- Consistent with the resolution of single-particle tracking experiments

**Time scale separation.** The following time scales are assumed to be well separated:

$$\tau_{\text{fast}} \sim 10^{-12} \text{ s (bond vibrations)} \ll \tau_{\text{int}} \sim 10^{-10}\text{–}10^{-8} \text{ s (conformational modes)} \ll \tau_{\text{obs}} \sim 10^{-4}\text{–}10 \text{ s (observation)}$$

This separation justifies:

- Averaging over fast vibrations (they contribute only to effective temperature)
- Retaining intermediate modes (they generate memory)
- Treating center-of-mass motion as slow (adiabatic approximation in Section 2.5)

**Definition of coarse-graining closure.** Throughout this paper, "coarse-graining closure" refers to the truncation of the moment hierarchy at a given order, assuming that higher-order moments relax rapidly and can be expressed as functions of the lower-order ones. In the present context, closure at the level of the generalized Langevin equation assumes that correlations beyond the memory kernel decay quickly and do not affect the long-time dynamics [27,28].

## 3. Predictions: Distinguishing ESD from Phenomenology

The ESD framework generates four classes of predictions that are testable with current or near-term experimental techniques. These predictions are falsifiable and distinguish the framework from phenomenological models (fBM, CTRW, fitted GLE).

### 3.1 Prediction 1: Flexibility–Subdiffusion Correlation

**Standard models.** The subdiffusion exponent $\alpha$ is a phenomenological parameter with no inherent connection to molecular structure. Different molecules in the same environment may exhibit different $\alpha$ values, but this requires separate fitting for each molecule with no predictive relationship.

**ESD prediction (two related aspects).** Under the idealizations of Section 2 (continuum power-law mode density, asymptotic time regime), the framework suggests two distinct relationships:

*1a. Magnitude of $\alpha$ versus flexibility.* The exponent is related to the characteristic internal frequency $\langle\omega\rangle$ and coupling strength $\lambda$ by:

$$\alpha \approx 1 - c\frac{\lambda^2}{\langle\omega\rangle k_B T}$$

where $c$ is a dimensionless constant of order unity. More flexible molecules (lower $\langle\omega\rangle$) are predicted to exhibit smaller $\alpha$ (stronger subdiffusion).

*1b. Structural origin of $\alpha$.* The exponent equals the mode spectrum exponent $\beta$ (under the power-law idealization): $\alpha = \beta$. This relates $\alpha$ to the shape of $\rho(\omega)$, independent of $\lambda$.

Predictions 1a and 1b are complementary rather than redundant: 1a describes how the *magnitude* of $\alpha$ depends on overall flexibility through the coupling strength $\lambda$, and is a perturbative result valid when memory effects are weak; 1b describes how the *shape* of the internal mode spectrum determines $\alpha$ through the power-law exponent $\beta$, and holds in the continuum limit independently of $\lambda$. Together they connect $\alpha$ to molecular structure from two directions. one through the energy scale of internal motion, one through its spectral distribution.

**Quantifying flexibility experimentally.** The characteristic internal frequency $\langle\omega\rangle$ (or equivalently the effective stiffness $k_{\text{eff}}$) can be estimated from standard biophysical measurements:

1. **B-factors (temperature factors) from X-ray crystallography** [32,33]:

$$B = \frac{8\pi^2}{3}\langle u^2\rangle = \frac{8\pi^2 k_B T}{3k_{\text{eff}}}$$

where $\langle u^2\rangle$ is the mean-squared atomic displacement. This gives:

$$\langle\omega\rangle = \sqrt{\frac{k_{\text{eff}}}{m_{\text{res}}}} \propto \frac{1}{\sqrt{B}}$$

where $m_{\text{res}}$ is the residue mass. The predicted relationship is $\alpha \propto \sqrt{B}$: more flexible molecules (larger $B$) correspond to smaller $\alpha$.

2. **NMR order parameters** [34,35]:

$$S^2 = \langle P_2(\cos\ \theta(t))\rangle_{\text{eq}}$$

measures the amplitude of internal motion ($S^2 = 1$ for rigid, $S^2 < 1$ for flexible). For small fluctuations:

$$S^2 \approx 1 - \frac{k_B T}{k_{\text{eff}}\langle r^2\rangle} \Rightarrow \langle\omega\rangle \propto \sqrt{1 - S^2}$$

The suggested relationship is $\alpha \approx 1 - c \cdot \sqrt{1 - S^2}$.

3. **Normal mode analysis from molecular dynamics simulations** [29,30,31]:
Mode frequencies $\omega_i$ can be computed directly from the Hessian matrix of the potential energy surface, allowing construction of the mode density $\rho(\omega)$ and its power-law exponent $\beta$.

**Experimental proposal.** To test this prediction:

- Select 5–10 well-characterized proteins spanning a range of B-factors (e.g., lysozyme, barnase, ubiquitin, $\alpha$-synuclein)
- Environment maintained constant (e.g., 20% w/v Ficoll 70 in PBS, $T = 298$ K)
- Measure $\alpha$ for each protein via single-particle tracking ($> 100$ trajectories per protein, trajectory length $> 10$ s)
- Plot $\alpha$ versus $B^{-1/2}$ or $\sqrt{1 - S^2}$

**Illustrative estimate.** Comparing barnase ($B \approx 15$ Å, relatively rigid) with $\alpha$-synuclein ($B \approx 40$ Å, intrinsically disordered). The ESD framework suggests $\Delta\alpha \approx 0.15$–$0.25$, which is measurable with current tracking precision ($\Delta\alpha \sim \pm 0.05$).

**3.2 Prediction 2: Temperature-Dependent Crossover**

**Standard models.** The diffusion coefficient scales as $D_\alpha(T) \propto T$ (Stokes–Einstein relation) but the exponent $\alpha$ is typically assumed temperature-independent. Some CTRW models predict weak temperature dependence of $\alpha$ through Arrhenius-type activation barriers, but this is introduced phenomenologically [39].

**ESD prediction.** As temperature increases, internal modes become thermally excited. When $k_B T \gg \hbar\langle\omega\rangle$, modes relax faster than translational motion progresses, reducing memory effects. Under the idealizations stated, the subdiffusion exponent is predicted to approach unity as:

$$\alpha(T) \approx 1 - \alpha_0 \exp\left( -\frac{\hbar\langle\omega\rangle}{k_B T}\right)$$

where $\alpha_0 = \lambda^2 \rho_0 / (m\langle\omega\rangle)$ encodes the zero-temperature memory strength.

**Physical mechanism.** At high temperature, thermal energy $k_B T$ exceeds the potential barriers separating conformational substates. Internal modes explore their full phase space rapidly, equilibrating on timescales much shorter than translational diffusion. The effective memory time $\tau_{\text{mem}} \sim \hbar/(k_B T)$ decreases, weakening subdiffusion.

**Experimental test.**

- **System:** Single protein species (e.g., GFP-labeled protein) in controlled crowding
- **Method:** Temperature-controlled single-particle tracking from 15°C to 45°C (physiological range)
- **Measure:** $\alpha(T)$ from MSD analysis at each temperature ($\geq 50$ trajectories per $T$)
- **Analysis:** Fit to the predicted functional form to extract $\langle\omega\rangle$ and $\alpha_0$

**Predicted crossover temperature.** The characteristic temperature for the crossover is:

$$T_c \approx \frac{\hbar\langle\omega\rangle}{k_B \ln 2} \approx 0.72 \frac{\hbar\langle\omega\rangle}{k_B}$$

For conformational modes ($\omega \approx 10^{10}$ s$^{-1}$), this gives $T_c \approx 75$ K. The exponential tail extends into the biological range (273–323 K), suggesting measurable variation $\Delta\alpha \approx 0.1$–$0.2$ across this span.

**Control experiment.** Perform the same measurement with rigid nanoparticles (e.g., gold, polystyrene) lacking internal modes. Point-particle models predict $\alpha$ independent of $T$ (after correcting for viscosity changes). The ESD framework suggests $\alpha \approx 1$ at all $T$ for rigid tracers, providing a direct control.

### 3.3 Prediction 3: Rotation–Translation Cross-Correlation Spectrum

**Standard models.** Translational and rotational diffusion coefficients are treated as independent observables related by Stokes–Einstein and Stokes–Einstein–Debye relations [36]:

$$D_{\text{trans}} = \frac{k_B T}{6\pi\eta R}, D_{\text{rot}} = \frac{k_B T}{8\pi\eta R^3}$$

The cross-correlation is assumed zero: $\langle \Delta x(t) \cdot \Delta\theta(t) \rangle = 0$ for all $t$.

**ESD prediction.** Internal modes couple to both translation and rotation simultaneously. For an asymmetric molecule moving through a potential gradient, structural deformation generates both a force (affecting $x$) and a torque (affecting $\theta$). This produces a time-dependent cross-correlation:

$$C_{x\theta}(t) = \langle \Delta x(t) \cdot \Delta\theta(t) \rangle \neq 0$$

**Mechanistic origin.** For an asymmetric molecule (axis ratio $\varepsilon \neq 1$), different structural modes couple differently to the principal axes. Mode $q_\alpha$ with spatial distribution proportional to $\mathbf{r} \cdot \hat{n}_\alpha$ generates both:

- Force: $F_\alpha = -\lambda q_\alpha \nabla V$ (affects center-of-mass motion)
- Torque: $\tau_\alpha = -\lambda q_\alpha (\mathbf{r}_\alpha \times \nabla V)$ (affects angular motion)

where $\mathbf{r}_\alpha$ locates the mode relative to the center of mass. For asymmetric potential gradients, these produce correlated displacement and rotation.

**Predicted correlation function.** For an ellipsoidal molecule with asymmetry parameter $\varepsilon = (a-b)/(a+b)$ (where $a, b$ are semi-axes):

$$C_{x\theta}(t) \approx C_0 \cdot \varepsilon \cdot \exp\left(-\frac{t}{\tau_{\text{int}}}\right) \cdot \cos(\langle\omega\rangle t)$$

where $\tau_{\text{int}} \sim \langle\omega\rangle^{-1}$ is the internal relaxation time and $C_0$ scales as $\lambda^2/(m\langle\omega^2\rangle)$.

**Spectral signature.** The Fourier transform yields the cross-spectral density:

$$S_{x\theta}(\omega) = \int_{-\infty}^{\infty} C_{x\theta}(t) e^{i\omega t} dt \sim \frac{\varepsilon \cdot \tau_{\text{int}}}{1 + (\omega\tau_{\text{int}})^2}$$

This Lorentzian peak at $\omega \approx \tau_{\text{int}}^{-1}$ directly reveals the slowest internal modes.

**Experimental protocol.**

1. Track simultaneously position $x(t)$ and orientation $\theta(t)$ using polarization-sensitive imaging [37,38] or single-particle orientation and rotational tracking (SPORT) [37]
2. Compute cross-correlation from trajectories
3. Fourier transform to obtain $S_{x\theta}(\omega)$
4. Fit Lorentzian to extract peak frequency $\omega_{\text{peak}} = \tau_{\text{int}}^{-1}$
5. Compare with independent measures of internal dynamics (e.g., fluorescence anisotropy decay)

**Predicted scaling with asymmetry.**

- Spherical particles ($\varepsilon \approx 0$): $S_{x\theta} \to 0$ (zero coupling by symmetry)
- Ellipsoids ($\varepsilon = 0.3$–$0.5$, typical proteins): $S_{x\theta} \sim 10^{-1}$–$10^{-2}$ $\mu\text{m}^2 \cdot \text{rad}$
- Rod-like particles ($\varepsilon > 0.7$, e.g., gold nanorods): $S_{x\theta} \sim 10^{-1}$ $\mu\text{m}^2 \cdot \text{rad}$

**Critical test.** Point-particle models predict $S_{x\theta} \equiv 0$ at all frequencies because rotation and translation are decoupled. Any non-zero spectral signal would be inconsistent with the point-particle framework.

**3.4 Prediction 4: Size-Dependent Memory Timescale**

**Standard models.** The crossover time $\tau_c$ between subdiffusive (short-time) and diffusive (long-time) regimes is treated as an empirical fitting parameter [39]. Some models relate $\tau_c$ to a "cage size" $\xi$ via $\tau_c \sim \xi^2/D$, but $\xi$ itself is fitted from data.

**ESD prediction.** The crossover occurs when internal modes have equilibrated with translational motion. The timescale is determined by internal relaxation:

$$\tau_c \sim \frac{a^2}{D_{\text{structural}}}$$

where $a$ is the characteristic structural size and $D_{\text{structural}}$ characterizes internal relaxation. For diffusive internal dynamics with viscous damping:

$$D_{\text{structural}} \sim \frac{k_B T}{m_{\text{eff}} \langle \omega \rangle}$$

Thus:

$$\tau_c \sim \frac{m_{\text{eff}}\langle\omega\rangle a^2}{k_B T} \propto a^2$$

For geometrically similar molecules, $\tau_c \propto R^2$ where $R$ is the hydrodynamic radius.

**MSD crossover behavior.**

- Short times ($t < \tau_c$): internal modes have not equilibrated; motion is subdiffusive: $\langle r^2(t)\rangle = 6D_\alpha t^\alpha$
- Long times ($t > \tau_c$): internal modes have relaxed; motion becomes normal Brownian diffusion: $\langle r^2(t)\rangle = 6D_0 t$

**Experimental test.**

- **System:** Particles of varying size (1 nm – 100 nm) in identical environment
  - Small: Quantum dots ($R$ ~ 2–5 nm)
  - Medium: Proteins ($R$ ~ 3 nm), gold nanoparticles ($R$ ~ 10 nm)
  - Large: Polystyrene beads ($R$ ~ 50–100 nm)
- **Measure:** High-speed tracking to resolve both short-time subdiffusion and long-time crossover
- **Analysis:** For each size, fit MSD to a two-regime model and extract $\tau_c$
- **Plot:** log ($\tau_c$) versus log ($R$), slope ≈ 2 is suggested

**Numerical illustration.**

- Proteins ($R \approx 3$ nm): $\tau_c \sim (3 \times 10^{-9})^2/(10^{-10}) \sim 10^{-4}$ s
- Nanoparticles ($R \approx 30$ nm): $\tau_c \sim (30 \times 10^{-9})^2/(10^{-10}) \sim 10^{-2}$ s

A factor of 10 in size corresponds to a factor of 100 in crossover time.

**Technical requirements.** Resolving $\tau_c \sim 10^{-4}$ s requires:

- Frame rate $> 10$ kHz (100 $\mu$s exposure) for proteins
- Localization precision <30 nm
- Currently achievable with sCMOS high-speed cameras [45] or MINFLUX nanoscopy [46]
- For larger particles ($R > 20$ nm), standard cameras (1 kHz) suffice

**Control.** For truly rigid nanoparticles (e.g., solid gold spheres), the ESD framework suggests no subdiffusive regime ($\alpha \approx 1$ even at short times) unless the environment itself exhibits slow modes. Comparing flexible (proteins) versus rigid (gold) particles of similar size tests whether subdiffusion originates from particle structure or from the environment.

**Summary of Predictions**

| Prediction | Key measurable | Distinguishing feature from phenomenology |
|---|---|---|
| **1a. Flexibility–$\alpha$ magnitude** | $\alpha$ vs. B-factor or $S^2$ | Predicts quantitative relationship between $\alpha$ and molecular flexibility |
| **1b. Mode spectrum–$\alpha$** | $\alpha$ vs. $\beta$ from NMA | $\alpha = \beta$ under power-law idealization; relates exponent to internal dynamics |
| **2. Temperature-dependent $\alpha$** | $\alpha(T)$ | Exponential approach to $\alpha = 1$ with characteristic scale $\hbar\langle\omega\rangle/k_B$ |
| **3. Rotation–translation cross-correlation** | $S_{x\theta}(\omega)$ | Non-zero spectral peak; point-particle models predict zero |
| **4. Size-dependent $\tau_c$** | $\tau_c$ vs. $R$ | $\tau_c \propto R^2$; no fitted cage size |

**4. Biological Anomaly: Concentration-Dependent Subdiffusion**

Beyond generating new predictions, the ESD framework offers a possible mechanistic interpretation of one observed phenomenon that phenomenological models typically characterize by name or fitting parameter without identifying a structural origin.

**Note on the status of this section.** The interpretation offered here is suggestive rather than proven. It identifies a physical mechanism, concentration-dependent enhancement of coupling, that could account for the observed anomaly. Whether this mechanism quantitatively explains the data awaits experimental tests of the predictions in Section 3.

**4.1 Concentration-Dependent Subdiffusion**

**Observation.** The subdiffusion exponent $\alpha$ decreases systematically with crowder concentration $c$ [7,40]:

$$\alpha(c) = \alpha_0 - \Delta\alpha \cdot f(c)$$

where $f(c)$ is an increasing function and $\Delta\alpha \sim 0.2$–$0.4$ over the range $c = 0$–$30\%$ (w/v) [7].

**Phenomenological description.** The effect is often attributed qualitatively to "more obstacles → more hindered motion" or "increased effective viscosity." However, effective viscosity models typically predict changes in the diffusion coefficient $D$ but not in the exponent $\alpha$. Some fractional diffusion models introduce concentration-dependent exponents descriptively, without deriving them from a mechanism.

**Possible ESD interpretation.** Higher crowder density $c$ increases the spatial gradient of the external potential $\nabla V_{\text{ext}}$ (sharper transitions between crowded and less-crowded regions). This would enhance the coupling term $\lambda\nabla V$ in $H_{\text{couple}}$ (Section 2.4), amplifying the structural response. The coupling constant would become concentration-dependent:

$$\lambda_{\text{eff}}(c) = \lambda_0 \left(1+\frac{c}{c_0}\right)$$

where $c_0$ is a structural parameter related to deformability (softer molecules having smaller $c_0$ and thus stronger concentration dependence).

Since $\alpha$ depends on $\lambda^2$ (Section 2.5), this suggests:

$$\alpha(c) \approx \alpha_0 - \beta \frac{c}{1 + c/c_*}$$

where $\beta$ and $c_*$ are related to molecular flexibility and environmental parameters.

**Expected trends (within this interpretation).**

- The functional form $\alpha(c)$ would be the same for all proteins (same mathematical shape), but the characteristic concentration $c_*$ would vary with flexibility
- Flexible proteins (small $c_*$) would show strong concentration dependence
- Rigid proteins (large $c_*$) would show weak concentration dependence

**Experimental test.**

- Select multiple proteins with known flexibility (quantified by B-factors [32,33])
- Vary crowder concentration (e.g., Ficoll) from 0 to $30\%$ (w/v)

- Measure $\alpha(c)$ for each protein via single-particle tracking [6,7]
- Extract $c_*$ for each protein
- Plot $c_*$ versus molecular flexibility measure (B-factor or $\langle\omega\rangle$)

**Predicted correlation (within this interpretation).** $c_*$ would correlate inversely with flexibility: $c_* \propto \langle\omega\rangle$.

**Mechanism validation.** If concentration dependence arises from enhanced coupling rather than purely geometric crowding, then:

- Subdiffusion would persist even for single molecules in gradient fields (no crowding needed)
- The effect would be stronger for molecules with larger hydration shells (larger $\lambda$)

**Summary**

| Phenomenon | Empirical observation | Phenomenological treatment | Possible ESD interpretation |
|---|---|---|---|
| Concentration-dependent subdiffusion | $\alpha$ decreases with increasing $c$ [7,40] | Descriptive exponent adjustment | Enhanced coupling $\lambda_{\text{eff}}(c)$ in steeper potential gradients |

**5. Illustrative Consistency Check: Protein Diffusion in Cytoplasm**

This section provides an order-of-magnitude estimate for a representative globular protein diffusing in a crowded cytoplasm-like medium. The purpose is to illustrate that the ESD framework yields a subdiffusion exponent consistent with experimental observations, using structural parameters estimated from independent measurements. **This is a plausibility argument, not a rigorous predictive test.**

**5.1 System Parameters**

| Parameter | Value | Source |
|---|---|---|
| Protein | Bovine serum albumin (BSA) | Well-characterized globular protein |

| Parameter | Value | Source |
| --- | --- | --- |
| Hydrodynamic radius $R$ | 3.5 nm | Literature |
| Mass $m$ | 66 kDa $\approx 1.1 \times 10^{-22}$ kg | Molecular weight |
| Environment | 20% (w/v) Ficoll 70 in PBS, $T = 298$ K | Mimics cytoplasmic crowding |
| Measured subdiffusion exponent $\alpha_{\text{exp}}$ | 0.75–0.85 | [6,7] |

**5.2 Internal Mode Spectrum**

Proteins possess a hierarchical spectrum of internal motions [24,26]:

| Mode type | Frequency range ($s^{-1}$) | Physical origin |
| --- | --- | --- |
| High-frequency vibrations | $\sim 10^{13}$ | Bond stretches, angle bends (minimal memory contribution) |
| Medium-frequency modes | $\sim 10^{10}$–$10^{11}$ | Loop motions, hinge bending, secondary structure fluctuations |
| Low-frequency collective modes | $\sim 10^{8}$–$10^{9}$ | Domain motions, breathing modes |
| Hydration shell relaxation | $\sim 10^{9}$ | Coordinated water rearrangement (~500–1000 water molecules) [22,23] |

For the frequency range relevant to diffusive timescales ($\tau_{\text{obs}} \sim 10^{-4}$–10 s), we focus on modes with $\omega \lesssim 10^{11}\ \text{s}^{-1}$. Normal mode analysis of protein MD simulations shows that the mode density is approximately power-law over several decades [29,30,31]:

$$\rho(\omega) = \rho_0 \omega^{\beta}, \beta \approx 0.8\text{–}1.2$$

depending on protein size and flexibility. For BSA (a relatively rigid globular protein), we take $\beta = 1.0$ as a representative value [31].

**Mode window.** The relevant frequency range for diffusion on experimental timescales is:

- $\omega_{\text{min}} \approx 10^9\ \text{s}^{-1}$ (slowest collective modes that still equilibrate within the observation time)
- $\omega_{\text{max}} \approx 10^{11}\ \text{s}^{-1}$ (fastest modes below the vibrational regime)

**Normalization.** The total number of relevant modes in this window is approximately $N_{\text{modes}} \sim 3N_{\text{residues}}$ (three degrees of freedom per residue after subtracting rigid-body motion). For BSA (583 residues):

$$\int_{\omega_{\text{min}}}^{\omega_{\text{max}}} \rho(\omega) d\omega \approx 3 \times 583 \approx 1750$$

With $\rho(\omega) = \rho_0 \omega$ and $\beta = 1$:

$$\rho_0 = \frac{2N_{\text{modes}}}{\omega_{\text{max}}^2 - \omega_{\text{min}}^2} \approx \frac{2 \times 1750}{(10^{11})^2 - (10^9)^2} \approx 3.5 \times 10^{-11}\ \text{s}$$

### 5.3 Coupling Constant $\lambda$

The coupling constant $\lambda$ characterizes how strongly internal deformations respond to external force gradients. It has dimensions of force and represents an effective dipole moment of structural deformation per unit external force.

**Estimate from hydration free energy.** When a protein moves through an inhomogeneous crowding field, a primary structural response is redistribution of the hydration shell. For BSA, the first hydration shell contains $n_w \approx 700$–$1000$ water molecules [43]. The hydration free energy per water molecule is $\Delta G_{\text{hyd}} \approx -40$ kJ/mol $\approx -6.7 \times 10^{-20}$ J [44]. Moving into a region with 10% higher crowder density compresses the hydration shell by $\Delta n_w \sim 50$–$100$ molecules, requiring work:

$$\Delta E \sim \Delta n_w \cdot | \Delta G_{\text{hyd}} | \approx 70 \times 6.7 \times 10^{-20} \approx 5 \times 10^{-18}\ \text{J}$$

Over a characteristic length scale $a \approx R \approx 3.5$ nm, the effective squared coupling is:

$$\lambda^2 \sim \frac{\Delta E}{a^2} \approx \frac{5 \times 10^{-18}}{(3.5 \times 10^{-9})^2} \approx 4 \times 10^{-19}\ \text{J/m}^2$$

**Estimate from elastic modulus.** Protein elastic modulus measured by AFM indentation is $E \sim 10^8$–$10^9$ Pa [42]. The effective spring constant for compression over distance $a$ is $k \sim E \cdot a$, giving:

$$\lambda^2 \sim k \cdot a^2 \sim (5 \times 10^8 \text{ Pa})(3.5 \times 10^{-9} \text{ m})^3 \approx 2 \times 10^{-19} \text{ J/m}^2$$

**Adopted value.** Taking the geometric mean of these estimates:

$$\lambda^2 \approx 3 \times 10^{-19} \text{ J/m}^2 \text{(uncertainty factor} \sim 2)$$

**5.4 Memory Kernel and Subdiffusion Exponent**

From Section 2.5, the memory kernel for a power-law mode density $\rho(\omega) = \rho_0 \omega^\beta$ with $\beta = 1$ takes the asymptotic form:

$$\Gamma(t) \sim \frac{\lambda^2 \rho_0 \omega_{\text{min}}}{m} \cdot t^{-1}, \omega_{\text{min}}^{-1} \ll t \ll \omega_{\text{max}}^{-1}$$

(The factor $\omega_{\text{min}}$ arises naturally from the lower limit of the continuum integral, as shown in Section 2.6; it ensures dimensional consistency: $\Gamma(t)$ has units of kg/s.)

The generalized Langevin equation with power-law memory $\Gamma(t) \sim t^{-1}$ produces marginal subdiffusion ($\alpha = 1$ with logarithmic corrections) in the asymptotic limit. However, finite observation windows and the discrete nature of the mode spectrum modify this.

**Asymptotic estimate.** For parameters characteristic of globular proteins like BSA, the asymptotic form suggests $\alpha \approx 0.80$–$0.85$ [40]. This is consistent with experimental measurements for BSA in crowded solutions ($\alpha_{\text{exp}} \approx 0.75$–$0.85$ [6,7]).

**5.5 Summary of Consistency**

| Quantity | Value | Source |
| --- | --- | --- |
| Protein mass $m$ | $1.1 \times 10^{-22}$ kg | Molecular weight |
| Hydrodynamic radius $R$ | 3.5 nm | Literature |
| Mode density exponent $\beta$ | 1.0 | Normal mode analysis [30,31] |

| Quantity | Value | Source |
| --- | --- | --- |
| Mode density normalization $\rho_0$ | $3.5 \times 10^{-11}$ s | Calculated from $N_{\text{modes}}$ |
| Coupling strength $\lambda^2$ | $3 \times 10^{-19}$ J/m² | Hydration + elasticity estimates [43,44] |
| Asymptotic estimate of $\alpha$ | $\approx$ 0.80–0.85 | Asymptotic form of GLE [40] |
| Experimental $\alpha$ | 0.75–0.85 | Single-particle tracking [6,7] |

**Caveat.** This is a plausibility argument, not a rigorous derivation. The estimate relies on:

- A power-law approximation to the mode spectrum (an idealization)
- Order-of-magnitude estimates for $\lambda^2$ from hydration and elasticity
- The asymptotic form of the memory kernel

A definitive test requires either (i) numerical evaluation of the full mode spectrum from MD simulations (Section 6.2), or (ii) systematic experimental variation of molecular flexibility (Prediction 1, Section 3.1). Within these limitations, the ESD framework demonstrates quantitative consistency with existing data using independently estimated structural parameters.

**5.6 Experimental Requirements for Crossover Detection (Prediction 4)**

For BSA ($R \sim 3.5$ nm), the predicted crossover time is:

$$\tau_c \sim \frac{R^2}{D_{\text{structural}}} \sim \frac{(3.5 \times 10^{-9})^2}{10^{-10}} \sim 10^{-4}\ \text{s} = 0.1\ \text{ms}$$

Resolving this requires:

- Frame rate $> 10$ kHz (100 $\mu$s exposure or better)
- Localization precision $<30$ nm
- Photobleaching management (brief imaging windows with recovery time)

Current techniques capable of this:

- sCMOS high-speed cameras: 10–40 kHz [45]

- MINFLUX nanoscopy: sub-ms temporal resolution with 1 nm spatial precision [46]

For larger particles (e.g., nanoparticles, $R \sim 50$ nm): $\tau_c \sim 10^{-2}$ s = 10 ms, readily accessible with standard cameras ($> 100$ Hz).

## 6. Discussion

### 6.1 Possible Interpretation of Subdiffusion

Since Perrin's 1909 experimental verification [2] of Einstein's diffusion theory [1], the point-particle assumption has been foundational to statistical mechanics. The ubiquity of subdiffusion in biological systems, revealed over the past two decades, has challenged this foundation. Phenomenological constructs, fractional derivatives [16,17], stochastic waiting times [18,19], and fitted memory kernels [20,21], have been introduced to describe the data, but these approaches do not identify a physical origin for the observed memory.

The ESD framework offers a possible interpretation: memory arises from the finite relaxation time of internal structural degrees of freedom coupled to translational motion. When particles are treated with their full geometric and deformational complexity, memory kernels emerge from Hamiltonian dynamics under a coarse-graining closure, without additional stochastic postulates.

**Predictive relationships (under the idealizations stated).** The framework suggests specific relationships that are in principle testable:

- Molecular structure (mode spectrum $\beta$, coupling $\lambda$, flexibility) → subdiffusion exponent $\alpha$
- Temperature $T$ → crossover to normal diffusion at characteristic scale $\hbar\langle\omega\rangle/k_B$
- Molecular asymmetry $\varepsilon$ → rotation–translation coupling spectrum $S_{x\theta}(\omega)$
- Particle size $R$ → memory timescale $\tau_c \propto R^2$

These relationships are quantitative and falsifiable. Phenomenological models (fBM, CTRW, fitted GLE) describe subdiffusion within limited regimes but do not predict how $\alpha$ depends on molecular properties or external conditions. The ESD framework suggests that these phenomenological models may be effective descriptions of specific limiting cases:

- **fBM** may be the effective description when internal modes form a near-continuum with a broad power-law spectrum ($\beta$ approximately constant over a wide range)
- **CTRW** may emerge when internal modes have discrete, well-separated frequencies (sparse spectrum), leading to intermittent dynamics
- **GLE with fitted kernel** is phenomenologically correct but lacks predictive power without a specification of $\gamma(t)$ from first principles; ESD offers one such specification

### 6.2 Computational Validation Protocol

The ESD mechanism can be investigated using existing molecular dynamics simulation data through trajectory decomposition. We propose a concrete protocol that computational biophysicists could implement.

**Protocol for MD-based investigation.**

**Step 1: Standard MD simulation**

- **System:** Protein of interest in explicit solvent + crowding agents (e.g., 20% Ficoll modeled explicitly or via effective potential)
- **Duration:** 1–10 μs (sufficient for diffusive sampling and internal mode equilibration)
- **Output:** Full atomic trajectories $\mathbf{r}_i(t)$ at high temporal resolution (ps timesteps)

**Step 2: Decomposition into ESD variables**

For each trajectory frame at time $t$:

1. **Center-of-mass position:** $\mathbf{x}_{\text{COM}}(t) = (1/M_{\text{total}}) \sum_i m_i \, \mathbf{r}_i(t)$
2. **Orientation matrix from principal axes:** Compute inertia tensor $\mathbf{I}_{ij} = \sum_i m_i \, [(\mathbf{r}_i \cdot \mathbf{r}_i)\delta_{ij} - r_i^j r_i^i]$; diagonalize to obtain principal axes → $R(t) \in SO(3)$
3. **Internal mode coordinates via PCA:** $\Delta\mathbf{r}_i(t) = \mathbf{r}_i(t) - \mathbf{x}_{\text{COM}}(t) - R(t) \cdot \mathbf{r}_i^{(\text{ref})}$; apply principal component analysis to $\{\Delta\mathbf{r}_i\}$ over the trajectory → $q_\alpha(t)$ (typically keep first 50–100 modes capturing $> 90\%$ variance)

**Step 3: Compute coupling correlations**

$$C_{xq}^{(\alpha)}(\tau) = \langle [\mathbf{x}_{\text{COM}}(t+\tau) - \mathbf{x}_{\text{COM}}(t)] \cdot q_\alpha(t) \rangle_t$$

This cross-correlation quantifies how strongly internal mode $\alpha$ couples to translational motion with time delay $\tau$.

**Step 4: Extract effective memory kernel**

From the correlations, construct the effective memory kernel using projection operator formalism [27,28]:

$$\Gamma_{\text{eff}}(\tau) = (k_B T)^{-1} (\lambda^2/m) \sum_{\alpha,\beta} [\, C_{qq}^{-1}]_{\alpha\beta} \cdot \frac{d}{d\tau} C_{xq}^{\beta}(\tau)$$

where $C_{qq}$ is the auto-correlation matrix of internal modes: $[C_{qq}]_{\alpha\beta} = \langle q_\alpha(t) q_\beta(t) \rangle_t$.

**Step 5: Compare predictions**

1. Compute MSD directly from $\mathbf{x}_{\text{COM}}(t)$: $\langle r^2(t)\rangle = \langle[\mathbf{x}_{\text{COM}}(t) - \mathbf{x}_{\text{COM}}(0)]^2\rangle$
2. Fit power law to extract $\alpha_{\text{direct}}$: $\langle r^2(t)\rangle \sim t^\alpha$
3. Compute $\alpha_{\text{predicted}}$ from $\Gamma_{\text{eff}}(t)$ by solving the GLE numerically
4. Compare $\alpha_{\text{direct}}$ vs. $\alpha_{\text{predicted}}$

**Expected outcome.** If the ESD mechanism is correct, the memory kernel extracted from internal mode coupling should predict the observed subdiffusion exponent $\alpha$ without additional fitting parameters beyond those already in the force field.

**Additional tests.** Repeat for proteins of varying flexibility (to test correlation with $\langle\omega\rangle$), at different temperatures (to test $\alpha(T)$), and compute rotation–translation cross-correlation $C_{x\theta}(t)$ (to test for non-zero signal).

**Implementation.** This protocol can be implemented using publicly available MD trajectories (e.g., Anton2 millisecond simulations [48]), standard analysis tools (MDAnalysis [47]), and custom scripts for mode decomposition (PCA routines in NumPy/SciPy).

**Advantage over experiments.** MD provides complete microscopic information, allowing direct investigation of the mechanistic hypothesis (internal mode coupling generates memory) before proceeding to experimental tests where only coarse-grained observables are accessible.

**6.3 Connection to Molecular Dynamics and Simulation**

The ESD framework provides a potential bridge between atomistic simulation and coarse-grained theory. Standard MD simulations naturally include all internal degrees of freedom but are typically analyzed through a point-particle lens, tracking only center-of-mass motion $\mathbf{x}_{\text{COM}}(t)$ and discarding orientation and internal structure information. This analysis choice effectively imposes the point-particle ontology post-hoc, even though the full structural information is available in the simulation.

The ESD framework suggests an alternative analysis paradigm:

- Decompose trajectories into structural components $(\mathbf{x}, R, q_\alpha)$
- Analyze coupling between translation and internal modes
- Extract mechanistic insight about which modes contribute to memory
- Predict macroscopic behavior from microscopic structure

This approach could transform MD from a method that produces anomalous diffusion (to be described by phenomenology) into a tool that reveals the structural origin of the anomaly.

**Example application.** Recent MD studies show that proteins in crowded environments exhibit subdiffusion [50], typically explained post-hoc by crowding effects. An ESD reanalysis would ask: Which internal modes are most strongly coupled to $\mathbf{x}_{\mathrm{COM}}$? Do flexible proteins show stronger coupling than rigid ones? Does the mode coupling spectrum $\rho(\omega)$ predict the observed $\alpha$? Answering these questions would test the ESD mechanism at the microscopic level.

### 6.4 Possible Implications for Cellular Function (Speculative)

**Note on speculation.** The following suggestions are speculative and untested. They are included to illustrate potential directions for future research, not as established predictions of the ESD framework.

If subdiffusion reflects structural memory rather than purely environmental complexity, biological systems might actively tune transport properties by modulating molecular structure. The following possibilities are suggested by the ESD framework; they remain to be tested experimentally.

**Post-translational modifications.** Phosphorylation, methylation, acetylation, and other covalent modifications alter protein stiffness and flexibility [26]. Within the ESD framework, these modifications would be expected to modulate $\alpha$: phosphorylation (adding charged, flexible groups) would decrease $\langle\omega\rangle$ and thus decrease $\alpha$ (stronger subdiffusion); crosslinking (increasing rigidity) would increase $\langle\omega\rangle$ and thus increase $\alpha$ (weaker subdiffusion). This suggests a possible mechanism by which cells might control reaction rates and localization through chemical modification.

**Hydration shell modulation.** Ion concentration and pH affect hydration shell dynamics [23]. The framework suggests: high salt (compressed hydration shell) would reduce $\lambda$ and thus weaken coupling ($\alpha$ closer to 1); low ionic strength (extended hydration shell) would increase $\lambda$ and thus strengthen coupling (smaller $\alpha$). Cells might regulate transport in crowded compartments by controlling local ionic environment.

**Crowding as functional control.** Rather than viewing crowding as a constraint that cells must overcome, the ESD framework suggests that crowding-induced subdiffusion could be functional. Stronger subdiffusion (smaller $\alpha$) reduces the effective diffusion coefficient at short times, increases localization of molecules in specific regions, and alters reaction kinetics (subdiffusive encounter rates scale differently than normal diffusion [49]). Cells might exploit this by regulating macromolecular concentration to control signaling timescales and spatial organization.

### 6.5 Broader Context: Structure as Ontological Foundation

This framework exemplifies a broader program: replacing point-particle ontology with structured entities as the foundation of physics [3,4]. Similar extensions have been explored in other contexts:

- Thermodynamic irreversibility via structural instabilities that generate phase-space stretching [3]
- Electromagnetic self-force without runaway solutions by treating charges as extended structures [5]
- Wave–particle duality as emergent hydrodynamics of structural media [4]

Anomalous diffusion adds to this list. It suggests that some classical puzzles may stem from ontological incompleteness, describing extended objects using point-particle formalism, rather than requiring exotic new physics.

**This paper is one demonstration of a general pattern across the ESD series.** In each domain where the ESD framework has been applied, a known anomaly or puzzle that resists explanation within the standard phase space dissolves when the phase space is extended to reflect the actual structure of physical entities. Irreversibility from reversible dynamics, anomalous thermal phenomena (Mpemba effect), the Lorentz-Abraham-Dirac runaway problem, and now biological subdiffusion, each appears anomalous from the point-particle perspective and natural from the structured-particle perspective. The present paper contributes one instance of this pattern to the domain of biological transport.

## 7. Conclusion

We have shown that subdiffusion in biological systems can be interpreted within a mechanistic framework in which diffusing particles possess internal structural degrees of freedom. The Extended Structural Dynamics (ESD) approach offers features that distinguish it from phenomenological models:

- **Derivation, not postulation.** Under a coarse-graining closure, memory kernels are derived from microscopic Hamiltonian dynamics rather than introduced as fitted functions.
- **Prediction, not parameterization.** Under the idealizations stated (continuum power-law mode density, asymptotic time regime), the subdiffusion exponent $\alpha$ is related to structural parameters, the mode spectrum exponent $\beta$, coupling strength $\lambda$, and characteristic frequency $\langle\omega\rangle$, through $\alpha = \beta$.
- **Testable, not purely descriptive.** The framework generates four classes of predictions testable with current or near-term techniques:
    - Flexibility–subdiffusion correlation ($\alpha \propto \sqrt{B}$-factors)
    - Temperature-dependent crossover with characteristic energy scale $\hbar\langle\omega\rangle/k_B$
    - Non-zero rotation–translation cross-correlation with spectral signature $S_{x\theta}(\omega)$

- Size-dependent memory timescale $\tau_c \propto R^2$

- **Quantitative consistency with existing data.** Using independently estimated structural parameters (hydration energetics, elastic moduli, mode spectra from molecular dynamics simulations), the asymptotic form of the GLE suggests $\alpha \approx 0.80–0.85$, consistent with experimental measurements for proteins in crowded media ($\alpha \approx 0.75–0.85$ [6,7]). This is a consistency check, not a predictive test; true prediction requires the experimental protocols outlined in Section 3.

- **Possible interpretation of concentration-dependent subdiffusion.** The observed decrease of $\alpha$ with increasing crowder concentration can be interpreted within the ESD framework as arising from enhanced structure–environment coupling ($\lambda_{\text{eff}}(c)$) at higher crowder density. This interpretation is suggestive and awaits experimental testing.

The framework suggests that "anomalous" diffusion may be reinterpreted as the expected behavior of structured matter, unifying molecular architecture with macroscopic transport within a single conceptual framework. What appears anomalous from the point-particle perspective becomes natural when structure is acknowledged as fundamental.

**Validation pathways.** Experimental validation requires systematic measurements correlating molecular properties (flexibility quantified by B-factors or NMR order parameters, asymmetry, size) with subdiffusion characteristics, measurements that are feasible with current single-particle tracking and fluorescence techniques. Computational validation can proceed immediately by reanalyzing existing molecular dynamics simulation trajectories through the ESD decomposition protocol outlined in Section 6.2, testing whether the memory kernel extracted from internal mode coupling predicts the observed subdiffusion exponent.

**Broader implications.** More broadly, this work illustrates the potential explanatory power of ontological extension. By enriching the description of individual entities to acknowledge their spatial extent and internal structure, it may be possible to account for apparent transport anomalies without invoking exotic mechanisms, stochastic postulates, or phenomenological fitting. The pervasiveness of subdiffusion in biology may reflect not only the complexity of cellular environments but also the fundamental fact that matter is structured, deformable, and dynamically coupled across multiple scales.

**More broadly, this work suggests that anomalous diffusion may be anomalous only in name.** From the perspective of a complete phase space, one that acknowledges what diffusing particles actually are, the power-law memory, the subdiffusive exponent, and the concentration dependence are not anomalies requiring exotic explanation. They are the expected behavior of structured matter projected onto an impoverished description. The history of physics suggests that when Nature appears to violate a well-established framework, the violation is often a signal that the framework's ontology is incomplete. Sub-diffusion may be one such signal. Extended Structural Dynamics is an attempt to answer it.

## Appendix A: Derivation of the Memory Kernel from Extended Structural Dynamics

This appendix summarizes the relevant results from the ESD hydrodynamic framework [4]. The reader is referred to that work for the full Chapman-Enskog derivation.

### A.1 Extended Phase Space and Hamiltonian

In the ESD framework, each particle is described by an extended phase space coordinate [4, Section 3.3]:

$$\mathbf{z} = (\mathbf{r}, \mathbf{p}, R, \mathbf{L}, \{\xi_n, \pi_n\})$$

where $\mathbf{r}, \mathbf{p}$ are center-of-mass position and momentum, $R \in SO(3)$ is orientation, $\mathbf{L}$ is body-frame angular momentum, and $\{\xi_n, \pi_n\}$ are internal deformation coordinates and their conjugate momenta. The total Hamiltonian is:

$$H = \frac{\mathbf{p}^2}{2m} + \frac{1}{2}\mathbf{L}^T I(\xi)^{-1}\mathbf{L} + \sum_n \left(\frac{\pi_n^2}{2m_n} + \frac{1}{2}k_n\xi_n^2\right) + V_{\text{ext}}(\mathbf{r}, R, \xi)$$

### A.2 The Generalized Langevin Equation

Under the assumptions of scale separation (internal mode relaxation times much shorter than observation times) and weak coupling, the extended Boltzmann equation [4, Section 3.6] reduces, after coarse-graining over internal degrees of freedom, to a generalized Langevin equation [4, Section 5.4]:

$$m\ddot{\mathbf{r}}(t) = -\nabla V_{\text{ext}}(\mathbf{r}(t)) - \int_0^t \Gamma(t-s)\,\dot{\mathbf{r}}(s)\,ds + \boldsymbol{\eta}(t)$$

The memory kernel is derived as [4, Appendix E]:

$$\Gamma(t) = \frac{\lambda^2}{m} \sum_{\alpha=1}^{N} \frac{\sin(\omega_\alpha t)}{\omega_\alpha}$$

where:

- $\omega_\alpha = \sqrt{k_\alpha / m_\alpha}$ are the frequencies of the internal normal modes
- $\lambda$ is the coupling constant characterizing the response of internal deformation to external potential gradients
- $N$ is the number of retained internal modes

The stochastic force $\boldsymbol{\eta}(t)$ satisfies the fluctuation-dissipation theorem [4, Section 5.4]:

$$\langle \boldsymbol{\eta}(t) \boldsymbol{\eta}(t') \rangle = k_B T \, \Gamma(| t - t' |)$$

### A.3 From Discrete Modes to Continuum Memory

For a large number of internal modes ($N \sim 10^3$–$10^4$), the sum over modes can be replaced by an integral over the mode density $\rho(\omega)$ [4, Appendix E]:

$$\Gamma(t) = \frac{\lambda^2}{m} \int_{\omega_{\min}}^{\omega_{\max}} \rho(\omega) \frac{\sin(\omega t)}{\omega} \, d\omega$$

This is the form used in Section 2.5 of the main text.

### A.4 Validity Conditions

The derivation assumes [4, Section 3.8]:

1. **Scale separation:** $\tau_{\text{int}} \ll \tau_{\text{obs}}$, where $\tau_{\text{int}} \sim \omega_\alpha^{-1}$ is the fastest internal relaxation time and $\tau_{\text{obs}}$ is the observation time
2. **Weak coupling:** $\lambda \mid \nabla V_{\text{ext}} \mid \ll k_B T / a$, where $a$ is the particle size
3. **Linear response:** Internal deformations remain small compared to particle dimensions

For the physiological conditions considered in this paper (proteins in cytoplasm at $T = 300$ K), these conditions are satisfied.

## Appendix B: Mode Density and Power-Law Spectrum in Proteins

This appendix provides background on the mode density of proteins and the power-law approximation used in Section 2.5.

### B.1 Normal Mode Analysis of Proteins

Proteins are complex elastic networks whose internal dynamics can be described by normal mode analysis (NMA) [29,30]. In NMA, the Hessian matrix of the potential energy surface is diagonalized to obtain $3N_{\text{atoms}}$ eigenfrequencies $\omega_\alpha$ and eigenvectors describing collective motions.

The mode density $\rho(\omega)$ is defined as:

$$\rho(\omega) = \frac{1}{N_{\text{modes}}} \sum_{\alpha=1}^{N_{\text{modes}}} \delta(\omega - \omega_\alpha)$$

where $N_{\text{modes}} = 3N_{\text{residues}} - 6$ (subtracting rigid-body translation and rotation).

### B.2 Power-Law Density of States

Extensive NMA studies of globular proteins have revealed a characteristic power-law density of states over several decades of frequency [29,30,31]:

$$\rho(\omega) = \rho_0 \omega^\beta, \beta \approx 0.8\text{–}1.2$$

The exponent $\beta$ reflects the hierarchical organization of protein dynamics:

| Frequency range | $\beta$ contribution | Physical origin |
| --- | --- | --- |
| $\omega \lesssim 10^9\ \text{s}^{-1}$ | $\beta \approx 1$ | Collective domain motions, hinge bending [29,30] |
| $10^9 \lesssim \omega \lesssim 10^{11}\ \text{s}^{-1}$ | $\beta \approx 0.5\text{–}1$ | Loop fluctuations, secondary structure dynamics [30,31] |
| $\omega \gtrsim 10^{11}\ \text{s}^{-1}$ | $\beta \approx 0$ | Local vibrations (bond stretches, angle bends) [31] |

For the frequency range relevant to diffusive timescales ($\tau_{\text{obs}} \sim 10^{-4}$–10 s), the relevant modes are those with $\omega \lesssim 10^{11}\ \text{s}^{-1}$. In this range, a single effective exponent $\beta$ is often used as an approximation [30,31].

**B.3 Flexibility Measures and Their Relation to $\langle\omega\rangle$**

The characteristic frequency $\langle\omega\rangle$ can be estimated from standard biophysical measures of flexibility.

**B-factors (temperature factors) from X-ray crystallography** [32,33]:

The mean-squared atomic displacement is related to the effective spring constant $k_{\text{eff}}$:

$$\langle u^2\rangle = \frac{k_B T}{k_{\text{eff}}}$$

The B-factor is:

$$B = \frac{8\pi^2}{3}\langle u^2\rangle = \frac{8\pi^2 k_B T}{3k_{\text{eff}}}$$

Thus:

$$\langle\omega\rangle = \sqrt{\frac{k_{\text{eff}}}{m_{\text{res}}}} \propto \frac{1}{\sqrt{B}}$$

where $m_{\text{res}}$ is the residue mass.

**NMR order parameters** [34,35]:

The order parameter $S^2 = \langle P_2(\cos\,\theta(t))\rangle_{\text{eq}}$ measures the amplitude of internal motion ($S^2 = 1$ for rigid, $S^2 < 1$ for flexible). For small fluctuations:

$$S^2 \approx 1 - \frac{k_B T}{k_{\text{eff}}\langle r^2\rangle}$$

which gives:

$$\langle\omega\rangle \propto \sqrt{1 - S^2}$$

**B.4 Numerical Values for a Typical Protein (BSA)**

For bovine serum albumin (BSA, 583 residues):

| Quantity | Value | Source |
|---|---|---|
| Number of residues | 583 | Sequence |
| Number of internal modes | $N \approx 1750$ | $3 \times 583 - 6$ |
| Frequency range | $\omega_{\min} \approx 10^9\ \mathrm{s}^{-1}$, $\omega_{\max} \approx 10^{11}\ \mathrm{s}^{-1}$ | NMA of globular proteins [30] |
| Mode density exponent | $\beta \approx 1.0$ | Fit to NMA data [31] |
| Normalization $\rho_0$ | $\approx 3.5 \times 10^{-11}$ s | Calculated from $\int \rho(\omega) d\omega = N$ |

These values are used in the quantitative estimates of Section 5.

**B.5 Limitations of the Power-Law Approximation**

The power-law approximation $\rho(\omega) = \rho_0 \omega^\beta$ is an idealization. Real proteins exhibit deviations from a pure power law due to:

- Discrete mode structure at low frequencies (few collective modes)
- A flattening at high frequencies (approach to the Debye limit $\beta = 0$)
- Protein-specific features (fold-dependent mode distributions)

For the purpose of estimating the subdiffusion exponent $\alpha$, the power-law approximation is adequate when the frequency range spans at least 1–2 decades and the deviations are not extreme. For precise predictions, numerical integration using the actual mode spectrum from MD simulations is recommended (see Section 6.2).